\magnification=1200

\font\it=cmti10

\font\tit=cmbx10

\font\ref=cmti9

\hoffset=1.6truecm
\voffset=1.0truecm
\hsize=6truein
\vsize=8.5truein
\baselineskip12pt
\centerline{\tit AN EXTENSION OF THE CAYLEY-HAMILTON THEOREM}
\centerline {\tit TO THE CASE OF  SUPERMATRICES}

\vskip.5cm

\baselineskip12pt
\centerline{L.F. Urrutia %
\footnote \dag { On sabbatical leave from Instituto de Ciencias  
Nucleares,
Universidad Nacional Aut\'onoma de M\'exico,
Circuito Exterior, C.U., 04510 M\'exico, D.F.} }
\centerline{Departamento de F\'\i sica}
\centerline{Universidad Aut\'onoma Metropolitana-Iztapalapa}
\centerline{Apartado Postal 55-534}
\centerline{09340 M\'exico, D.F.}

\centerline{and}

\centerline{Centro de Estudios Cient\'\i ficos de Santiago}
\centerline{Casilla 16443, Santiago 9}
\centerline{Chile}

\

\centerline{and}

\

\centerline{N. Morales}
\centerline{Departamento de F\'\i sica }
\centerline{Universidad Aut\'onoma Metropolitana-Iztapalapa}
\centerline{Apartado Postal 55-534}
\centerline{09340 M\'exico, D.F.}
\vskip 1cm
\baselineskip20pt

\centerline{\tit ABSTRACT}

Starting from the expression for the superdeterminant of
$ (xI-M)$, where
$M$  is an arbitrary supermatrix , we propose a definition for the
corresponding
characteristic polynomial and we prove that each supermatrix
satisfies its  characteristic equation.
Depending upon the factorization properties of the  basic polynomials
whose ratio defines  the above mentioned superdeterminant we are
able to construct polynomials of lower degree which are also shown to be
annihilated
by the supermatrix.


\vfill\eject



\noindent{1.-} INTRODUCTION

Given any $n \times n$ real matrix $M$, its characteristic
polynomial is defined by $P(x) = det (xI - M)$, where $I$ denotes
the $n \times n$ identity matrix and $x$ is a real variable. In
general $P(x) = x^n + \sum_{k=0}^{n-1} c_k x^k$ is a monic
polynomial of degree $n$. The Cayley-Hamilton theorem asserts that
$P(x = M) = 0$. That is to say, if we substitute in $P(x)$ the
real variable $x$ by the matrix $M$ in all the powers $x^k (k \not=
0)$, and set $x^0 = I$, we obtain the matrix zero as the result.
The coefficients $c_k (k
\not= 0)$ can be written in terms of $Tr(M), Tr(M^2),\cdots ,
Tr(M^{n-1})$ together with their powers and $c_0 = det (M)$.
This theorem has recently found interesting applications
in 2+1 dimensional Chern-Simons  (CS) theories [1]. Pure CS
theories are of topological nature and the fundamental degrees
of  freedom are the traces of group elements constructed
as the holonomies ( or Wilson lines, or integrated  connections)
of the gauge connection around oriented closed curves on the
manifold. The observables are the expectation values of the Wilson  
lines which
turned out to be realized as the various knot polynomials
known to  mathematicians [2]. Since CS theories are also exactly
soluble and possess a finite number of degrees of freedom [3], another
aspect of interest is the reduction of the initially  
infinite-dimensional phase
space to the subspace of the true degrees of
freedom. The Cayley-Hamilton theorem has played an important
role in the construction of the so called skein  relations [4],
which are relevant to the calculation of expectation values, and
also in the process of reduction of the phase space. To illustrate
the basic ideas related to this last point let us consider
the simple case of
two matrices $M_1$ and $M_2$ which
belong to $SL(2,R)$. In this case the characteristic polinomial
is $P(x) = x^2 - Tr(M_1) x + 1$ and we have the
Cayley-Hamilton matrix identity
$$(M_1)^2 - Tr(M_1) M_1 + I = 0. \eqno(1.1)$$



\noindent
Multiplying Eq. (1.1) by $M_2 M_1^{-1}$ and taking the trace we  
obtain the
following non-linear constraint among the traces
$$Tr(M_2 M_1^{-1}) + Tr(M_1 M_2) = Tr(M_1) Tr(M_2).
\eqno(1.2)$$

\noindent
The expression (1.2) finds a very useful application
in the discussion of the reduced phase space of the de Sitter
gravity in 2 + 1 dimensions, which is equivalent to the
Chern-Simons theory of the group $SO(2,2)$ [3]. This theory can be
more easily described in terms of two copies of the group
$SL(2,R)$, which is the spinorial group of $SO(2,2)$. The gauge
invariant degrees of freedom associated to one genus of an
arbitrary genus $g$ two-dimensional surface turn out to be traces
of any product of powers of two $SL(2,R)$ matrices $M_1$ and
$M_2$, which correspond to the holonomies (or integrated
connections) of the two basic  homotopically distinct trajectories on
one genus. Nevertheless, because Chern-Simons theories have a
finite number of degrees of freedom, one should be able to reduce
this infinite set of traces to a finite one. This task can in
fact be accomplished by virtue of the relation (1.2). In other
words,  $Tr(M_1{}^{p_1} M_2{}^{q_1} M_1{}^{p_2}
M_2{}^{q_2} \cdots M_1{}^{p_n} M_2{}^{q_n} \cdots )$, for any
$p_i, q_i$ in $\cal{Z}$, can be shown to be reducible and to be expressed
as a function of three traces only: $Tr(M_1), Tr(M_2)$ and
$Tr(M_1 M_2)$ [5].
A similar reduction can be performed in the case of 2 + 1 super
de Sitter gravity, which is the Chern-Simons theory of the
supergroup $Osp(2\vert 1, C\hskip-7pt I \ )$ [6]. The novelty here  
is that one
is
dealing with supermatrices instead of ordinary matrices. In the
particular case considered, a Cayley-Hamilton identity for the
supermatrices was obtained in an heuristical way and a relation
analogous to (1.2) was derived. This allowed to carry out the
reduction of the infinite dimensional phase space in the one-genus  
sector of
the theory, this time in
terms of five complex supertraces [7]. We observe that the  non-linear
constraints among the traces that need to be solved in order to  
acomplish the
reduction of the phase space, of which Eq.(1.2) is an example, are  
ussually
obtained starting from the so called Mandelstam identities [8]. The  
 discussion
of the relation between the  Cayley-Hamilton  and the Mandelstam   
identities,
together with the construction of the latter identities in the case of
supermatrices is reported in Ref.[9]. It is important to emphasize  
also that
the use  of the Mandelstam identities is of fundamental importance in the
formulation of arbitrary gauge theories in terms of Wilson loops  
variables,
which constitute an overcomplete set of degrees of freedom [10].

In this paper we discuss the general construction of
Cayley-Hamilton type identities for supermatrices. This is an
interesting problem in its own, besides  possible applications like  
:  (i)
 the study of the reduced  phase space in Chern-Simons theories
defined over a supergroup or (ii)  the loop space formulation of  any
supersymmetric gauge theory . In Section 2 we introduce our notation
and  we  propose a definition of
the characteristic and null polynomials for supermatrices, starting  
from the
corresponding superdeterminant. In Section 3 we prove the
Cayley-Hamilton theorem for the polynomials
previously defined.

\vskip 1.4pc
\noindent
2.- THE CHARACTERISTIC AND NULL POLYNOMIALS FOR SUPERMATRICES

We consider a Grassmann algebra $\Lambda={\Lambda}_{0}  \oplus  
{\Lambda}_{1}$
over the complex numbers
$C\hskip-7pt I \ $, where $ {\Lambda}_{0}$ (${\Lambda}_{1}$) is the  
even (odd)
part of $ \Lambda$. Any element $a \in \Lambda$ is a sum of the
body $\bar a \in C \hskip-7pt I \ $ plus the nilpotent element
$s(a)$ called the soul. The ring of polynomials over this
Grassmann algebra is denoted by $\Lambda_0[x]$ and consists of all
polynomials
$f(x) = a_0 x^n + a_1 x^{n-1} + \cdots + a_n $,
where $a_k$ are even elements of the Grassmann algebra.  The  
Grassmann algebra
$\Lambda$ is generated by an infinite number of  odd generators $  
{\xi}^{A}$.
Nevertheless, when dealing with an specific supermatrix  we consider only
superfunctions of the given supermatrix elements. These elements  
will have  an
expansion in terms of the basis $ \left\{ {\xi}^{A} \right\}$,  
which is not
relevant for our purposes [11].

A $(p+q) \times (p+q)$ supermatrix is a block matrix of the form
$$M= \left( \matrix{A & B\cr
C & D\cr}
\right), \eqno(2.1)$$

\noindent
where $A, B, C$ and $D$ are $p\times p, p \times q, q\times p,
q\times q$ matrices respectively. The distinguishing feature with
respect to an ordinary matrix is that the matrix elements
$M_{RS} \ R= (i, \alpha), \ S= (j,\beta)$ are elements of $ \Lambda$
 with the property that $A_{ij} \ (i,j = 1, \cdots p)$ and
$D_{\alpha\beta} \ (\alpha,\beta = 1, \cdots q)$ are even
elements, while $B_{i\alpha}$ and $C_{\beta j}$ are odd elements
of the algebra. In particular this means that such numbers satisfy
$$\eqalign{B_{i\alpha} B_{j\beta} = -&B_{j\beta} B_{i\alpha}, \   
C_{\alpha i}
C_{\beta j} = - C_{\beta j} C_{\alpha i} \cr
&B_{i\alpha} C_{\beta j} = -C_{\beta j} B_{i\alpha}, \cr}\eqno(2.2)$$

\noindent
while $A_{ij}$ and  $D_{\alpha\beta}$
commute with everything.

Let us recall that the ordinary matrix addition and the ordinary
matrix product of two supermatrices is again a supermatrix.
Nevertheless, such concepts as the trace and the determinant need
to be redefined, because of the odd component piece of the
supermatrix.

The basic invariant under similarity transformations for
supermatrices is the supertrace, defined by
$$Str(M) = Tr(A) - Tr(D), \eqno(2.3)$$

\noindent
where the trace ($Tr$) over the even matrices is the standard one.  An
important property of the above definition is the cyclic identity
$Str(M_1M_2) = Str (M_2M_1)$, for arbitrary supermatrices, which
is just a consequence of the relative minus sign in (2.3). The
generalization of the determinant, called the superdeterminant ($Sdet$),
is obtained from (2.3) by defining
$$ Sdet (M )= \exp Str (ln M), \eqno(2.4)$$

\noindent
which leads to the  the following
equivalent  expressions for  the superdeterminant [12]
$$Sdet (M) = {det(A-B D^{-1} C)\over det D} = {det A \over
det(D-C A^{-1} B)}. \eqno(2.5)$$

\noindent
All the matrices involved now are even in the Grassmann
algebra and the determinant ($det$) has its usual meaning. The  
superdeterminant
inherits the basic property  $Sdet(M_1 M_2)$ = $Sdet (M_2 M_1)$ and  
requires
$det D \not= 0 $
and $det A \not= 0$ in order to be defined. An explicit
demonstration of the equality of the  alternative ways (2.5) of
calculating $Sdet (M)$ is given in Ref. [13].

In order to proceed we introduce $a(x) = det (xI-A)$ and $d(x) =
det (xI-D)$, which are the characteristic polynomials of the even
matrices $A$ and $D$.

Starting from the two alternative expressions  (2.5) of calculating the
superdeterminant we find it convenient to state the following:

{\bf Lemma 2.1} \ \  For any $(p+q) \times (p+q) $ supermatrix $M$,   the
characteristic function $h(x) = Sdet (xI-M)$ can be written as
$$h(x) = {\tilde F(x)\over \tilde G(x)} = {F(x) \over G (x)},
\eqno(2.6)$$
\noindent
where the basic polynomials $\tilde F$, $\tilde G$, $F$ and $G$ are  
given by
$$\tilde F(x) = det (d(x) (xI-A) - B adj(xI-D) C), \ \ \tilde
G(x) = (d(x))^{p+1}, \eqno(2.7a)$$
$$F(x) = (a(x))^{q+1}, \ \ G(x) = det (a(x) (xI-D) - C adj
(xI-A)B). \eqno(2.7b)$$

{\bf Proof.} The above expressions are directly obtained from  
Eqs.(2.5)  using
the  relation $(xI-F)^{-1} = [det(xI-F)]^{-1} adj
(xI-F)$ valid for any even matrix $F$. Notice that $\tilde F$ is
expressed in terms of the determinant of a $p \times p$ even
matrix, while $G(x)$ is the determinant of a $q \times q$ even
matrix.

In order to motivate the basic idea of our definition for the
characteristic polynomial of a supermatrix let us consider the
simple case of a block-diagonal supermatrix $M \ (i.e. \ B = 0, C =
0)$. Here $h(x) = a(x)/d(x)$ and clearly the characteristic
polynomial is $P(x) = a(x) d(x)$, which is the product of the  
numerator and the
denominator of the corresponding  superdeterminant. In fact we have
$$P(M) = \left( \matrix{ a(A) & 0\cr
0 & a(D) \cr}\right) \
\left(\matrix{ d(A) & 0\cr
0 & d(D)\cr} \right) \equiv 0 \eqno(2.8)$$

\noindent
because $a(A) = 0, d(D) = 0$. In the general case where $h(x)$ is  
given by
Eq.(2.6), the numerator of the superdeterminant is  $\tilde F$  \   
($F$)  while
the denominator is  \  $ \tilde G$  ($G$), which motivates the following:

{\bf Definition 2.1} \ \ For an arbitrary $(p+q) \times (p+q) $  
supermatrix $M$
we define the  characteristic polynomial
$${\cal P}(x) = \tilde F (x) G(x) = F(x) \tilde G(x),  \eqno(2.9)$$
where the basic polynomials $\tilde F, \tilde G, F, $ and $G$ are  
given in
Eqs.(2.7). For notational simplicity we will not necessarily write  
explicitly
the $x$-dependence on many of the polynomials considered in the sequel.

 When $a(x)$ and
$d(x)$ have a common factor $f(x)$ in the block-diagonal case,
$$a(x) = f(x) a_1(x) , \ d(x) = f(x) d_1 (x), \eqno(2.10)$$

\noindent
the characteristic polynomial is given by $P(x) = f(x) a_1 (x) d_1(x),$
which is a polynomial of lower degree than the product $a(x)
d(x)$.
Motivated by this fact together with the work of Ref. [14], we have
realized that there are some cases in which we can construct null
polynomials of lower degree than ${\cal P}(x)$, according to the  
factorization
properties of the basic polynomials $\tilde F, \tilde G, F, G$.
At this point it is important to observe that we do not have a
unique factorization theorem for polynomials defined over a
Grassmann algebra. This can be seen,  for example,  from the  
identity $x^2 =
(x+\zeta\alpha)(x-\zeta\alpha)$, where $\alpha$ is an even Grassmann with
$\alpha^2 = 0$ and $ \zeta$ is an arbitrary complex number. The  
construction of
the null polynomials of lower degree starts
from finding the divisors of maximun degree of the pairs $\tilde
F, \tilde G, (F,G)$ which we denote by $R(S)$ respectively. This
means that one is able to write
$$\eqalign{\tilde F = R \tilde f, \ \  \tilde G = R \tilde g, \cr
F = Sf, \ \  G = Sg,\cr}\eqno(2.11)$$

\noindent
where all polynomials are monic and also $\tilde f, \tilde g, f,
g$ are of least degree by construction. They must satisfy
$$\tilde f/\tilde g = f/g,  \eqno(2.12)$$

\noindent
because of Eq. (2.6) and the expressions in (2.11) might be
not unique. Let us observe that in the case of polynomials over
the complex numbers Eq. (2.12) would imply at most $\tilde f =
\lambda f, \tilde g = \lambda g$ with $\lambda$ being a constant.
Since we are considering polynomials over a Grassmann algebra
this is not necessarily true as can be seen again in the above
mentioned identity $x/(x- \zeta\alpha) = (x+\zeta\alpha)/x$, which  
we have
rewritten in a convenient way. The above discussion leads  to the  
following:

{\bf Definition 2.2}\ \ Given an arbitrary $(p+q) \times (p+q)$  
supermatrix
$M$, with a characteristic function $h(x)$  such that $\tilde F,  
\tilde G$
have a common factor   $R$ ($\tilde F=R \tilde f, \   \tilde G=R  
\tilde g ) $
and $ F, G$ have a common factor $S$ ($ F=S f, \  G=S g ) $, where  
$ \tilde
f/\tilde g=f/g, $ we define  a null polynomial of $M$  by
$$P(x) = \tilde f(x) g (x) = f(x) \tilde g (x).  \eqno(2.13)$$

The above polynomial is clearly of lower degree than ${\cal P}(x)$,  
  which is
just a particular case
of the null polynomials (2.13)  when $R=S=1$.  We will
concentrate mostly on Def. (2.2) in the sequel.

\vskip 1.4pc
\noindent
3.- THE CAYLEY-HAMILTON THEOREM FOR SUPERMATRICES

In this section we prove that the polynomial given in Def.(2.2)  
does  in fact
annihilate the supermatrix $M$. The first step of our strategy to  
prove  the
Cayley-Hamilton theorem for supermatrices
 is based on one of the
standard methods to prove such theorem for
ordinary matrices [15]. We briefly recall such procedure and
emphasize that it is independent of the matrix considered being a
standard matrix or a supermatrix.

{\bf Lemma 3.1}. Let $M$, $(xI-M)$ and $N(x)$ be $(p+q)\times (p+q)$
supermatrices where $M$ is independent of $x \in \Lambda_0$ , with $N(x)$
being a polynomial supermatrix  of degree $(n-1)$,  $N(x) = N_0x^{n-1}
+ N_1 x^{n-2} + .. + N_{n-1} x^0$, (where each $N_k \ (k = 0,
\cdots , n-1)$ is a $(p+q)\times(p+q)$ supermatrix independent of  
$x$ ) such
that
$$(xI-M) N(x) = P(x) I, \eqno (3.1)$$
where $P(x) = p_0 x^n + p_1 x^{n-1} + \cdots + p_n x^0$ is a
numerical polynomial of degree $n$  $\in \Lambda_0[x]$, then $P(M)  
= p_0 M^n +
p_1
M^{n -1} + \cdots + p_n I \equiv 0$.

{\bf Proof}.  The proof follows by
comparing the independent powers of $x$ in Eq. (3.1) and then explicitly
computing $P(M)$ [15].

In the standard case the polynomial matrix
$N(x)$ is just given by $N(x) = adj (xI-M) = det(xI-M)
(xI-M)^{-1}$, and $P(x) = det (xI-M)$. In the case of a supermatrix  
we do not
have an obvious
generalization either of the matrix $adj (xI-M)$ or
of $det (xI-M)$. Nevertheless, following the analogy as close as
possible we define
$$N(x) = P(x) (xI-M)^{-1}, \eqno(3.2)$$

\noindent
where $P(x)$ is the polynomial
introduced in Def.(2.2) of the previous section. The challenge now  
is to prove
that $N(x)$, which trivially satisfies the Eq. (3.1), is indeed a
polynomial matrix. In this way we would have proved that $P(M) =
0$, according to Lemma 3.1. To this end we consider the following :

{\bf Lemma 3.2.}  Let $M$ and $(xI-M)$ be $(p+q)\times (p+q)$  
supermatrices, $x
\in \Lambda_0$, then
$$(xI-M)^{-1}_{ij} = - {1\over \tilde F}{\partial \tilde F\over
\partial A_{ji}},\ \ \ \ \ (xI-M)^{-1}_{i \alpha } =  {1\over G}  
{\partial
G\over \partial C_{\alpha i}} \eqno(3.3a)$$
$$
(xI-M)^{-1}_{\alpha j} =  {1\over \tilde F} {\partial \tilde F\over
\partial B_{j \alpha }}, \ \ \ \ \ (xI-M)^{-1}_{\alpha\beta} = -  
{1\over G}
{\partial G\over
\partial D_{\beta\alpha}},\eqno(3.3b)$$
where $A_{ij},B_{j  \alpha},C_{\alpha j} $and $D_{\alpha \beta}$
are the entries of the supermatrix  $M$ defined in Eq. (2.1)
and  $\tilde F$, \  $G$,  are the polynomials given in Eqs. (2.7). The
derivative with respect to an odd Grassmann number is  a left derivative
defined such that  $\delta\tilde F
\equiv \delta B_{j\alpha} {\partial\tilde F\over \partial
B_{j\alpha}}$.

{\bf Proof. }  The first step is to calculate $(xI-M)^{-1}$ in  
block form, with
the results
$$(xI-M)^{-1}_{11} = ((xI-A)-B (xI-D)^{-1} C)^{-1}, \eqno(3.4a)$$
$$(xI-M)^{-1}_{12} = -(xI-A)^{-1}B ((xI-D) - C(xI-A)^{-1}  
B)^{-1},\eqno(3.4b)$$
$$(xI-M)^{-1}_{21} = -(xI-D)^{-1}C ((xI-A)-B (xI-D)^{-1}C)^{-1},
\eqno(3.4c)$$
$$(xI-M)^{-1}_{22} = ((xI-D)-C (xI-A)^{-1} B)^{-1}, \eqno(3.4d)$$

\noindent
where the subindices 11, 12, 21 and 22 denote the corresponding
$p\times p, p\times q, q\times p$, and $q\times q$ blocks. Let us  
concentrate
now in the $11$ block. Rewritting all the inverse matrices in  
Eq.(3.4a) in
terms of their adjoints together with the corresponding  
determinants we obtain
$$(xI-M)^{-1}_{11}= {d\over \tilde F} adj ( (xI-A)d - Badj
(xI-D) C). \eqno(3.5)$$
Using the basic property
$$\delta det Q = Tr (adj Q \delta Q),\eqno(3.6)$$
\noindent
valid for any even matrix $Q$, we calculate the change of $\tilde F$ with
respect to $A_{ij}$, keeping constant all other entries, obtaining
$$\delta\tilde F = - d \left[ adj ((xI-A)d - B adj
(xI-D)C\right]_{ij} \ \delta A_{ji}, \eqno(3.7)$$

\noindent
which can be written as
$${\partial\tilde F\over \partial A_{ji}} = - d
\left[ adj ((xI-A)d - B adj (xI-D) C)\right]_{ij}. \eqno(3.8)$$

\noindent
The comparison of Eq.(3.8) with Eq. (3.5) completes the proof of  
the first
relation in Eq. (3.3a). The corresponding proof for the remaining  
Eqs. (3.3) is
performed following a similar procedure.

We observe  that the conditions for the existence of
$(xI-M)^{-1}$ are the  same as those for the existence of $Sdet
(xI-M)$ and they are $det(xI-A) \not= 0$ and $det(xI-D) \not=
0$. Since $x$ is a generic even Grassmann variable we will assume that
this is always the case. By virtue of these assumptions the term
$((xI-A) - B(xI-D)^{-1} C)^{-1}$, for example, can always be  
calculated as
$(I-(xI-A)^{-1}B (xI-D^{-1}) C)^{-1} (xI-A)^{-1}$. The factor on the
left can be thought as a series expansion of the form $1/(1-z) = 1  
+z+z^2+
\cdots ,$ with $z= (xI-A)^{-1}B
(xI-D)^{-1} C$. Moreover, the series will stop at some power
because $z$ is a matrix with body zero and thus it is nilpotent.

Now we come to the principal result of this paper, which we state as the
following:

{\bf Theorem 3.1.} Let $M$ and $(xI-M)$ be $(p+q)\times (p+q)$  
supermatrices,
$x \in \Lambda_0$, then $N(x) = P(x) (xI-M)^{-1}$, with $P(x)$ given in
Def.(2.2), is a polynomial matrix.

{\bf Proof.} Let us consider the block-element 11  of
$N(x)$ to begin with. According to Lemma (3.2)
 together with Eq. (2.11), this block can be written as
$$N_{ij} = - g {\partial \tilde f\over \partial A_{ji}} -
{g \tilde f \over R} {\partial R\over \partial A_{ji}}.
\eqno(3.9)$$

\noindent
The first term of the RHS is clearly of polynomial character. In
order to transform the second term we make use of the property
$${\partial ln \tilde G\over \partial A_{ji}} = 0 = {\partial ln
R\over \partial A_{ji}} + {\partial ln \tilde g \over \partial
A_{ji}}, \eqno(3.10)$$

\noindent
which follows from the factorization $\tilde G = R \tilde g$,
together with the fact that $\tilde G$ is just a function of
$D_{\alpha\beta}$, according to Eq. (2.7a). In this way, and using   
also the
Eq.(2.12),  we obtain
$$N_{ij} = f {\partial \tilde g\over \partial A_{ji}} -
g {\partial \tilde f\over \partial A_{ji}}, \eqno(3.11)$$

\noindent
which leads to the conclusion that the block-matrix $N_{ij}$ is
indeed polynomial. The proof for $N_{\alpha i}$ runs along the
same lines, except that now the derivatives are taken with
respect to $B_{i\alpha}$ and that we have to use ${\partial ln
\tilde G\over \partial B_{i\alpha}}= 0$, instead of Eq. (3.10).
The remaining terms $N_{i\alpha}$ and $N_{\alpha\beta}$ can be
dealt with in analogous manner by considering the derivatives of
$ G =S g$ with respect to $C_{\alpha i}$ and
$D_{\beta\alpha}$, and by replacing the condition (3.10) by
${\partial ln F\over \partial C_{\alpha i}} = 0 $ and ${\partial
ln F\over \partial D_{\beta\alpha}} = 0 $ respectively. The
results are again of the form (3.11), the only difference been the
variables with respect to which the derivatives are taken.

Finally, using Theorem (3.1) together with Lemma (3.1)  we can state the
following extension of the Cayley-Hamilton theorem to the case  of
supermatrices:

{\bf Theorem 3.2.} (Extended Cayley-Hamilton Theorem)  Let $M$ and  
$(xI-M)$ be
$(p+q)\times (p+q)$ supermatrices, $x \in \Lambda_0$, with  
$Sdet(xI-M)=\tilde
F/\tilde G=F/G$,  where the polynomials $\tilde F, \tilde G, F$ and  
$G$ are
given in Eqs.(2.7). Then, for any  common factor $R$ such that  
$\tilde F = R
\tilde f, \  \tilde G = R \tilde g$ and $S$ such that $F = Sf, \  G  
= Sg$,
where $\tilde f/\tilde g = f/g,  $ the polynomial $P(x)=\tilde f(x)  
g (x) =
f(x) \tilde g (x)$  annihilates $M$, i.e.  $P(M)=0.$

A less formal presentation of the above results can be found in  
Refs.[16]. A
detailed version of this work containing many examples is given in  
Ref. [17].

\vfill
\eject

\noindent
ACKNOWLEDGEMENTS

The work of both authors has been partially supported by the grant
DGAPA-UNAM-100691. LFU also  acknowledges support from the grant
CONACyT-0758-E9109.

\

\noindent
REFERENCES

\

\item{1.} For a review see for example Birmingham, D., Blau, M.,
Rakowski, M. and Thompson, G., {\it Phys. Rep.} {\tit 209}, 129 (1991).

\item{2.} Jones, V., {\it Bull AMS} {\tit 12}, 103 (1985) and
{\it Pacific J. Math.} {\tit 137}, 312 (1989).

\item{ \ } Freyd, P., Yetter, D., Hoste, J., Lickorish, W.,
Millet, K. and Ocneanu, A.,  {\it Bull AMS} {\tit 12}, 239 (1985).

\item{ \ } Kauffman, L.,  {\it Topology} {\tit 26}, 395 (1987).

\item{ \ }Witten, E.,  {\it Commun. Math. Phys.}
{\tit 121}, 351 (1989)
and {\it Nucl. Phys.} {\tit B322}, 629 (1989 ).

\item{3.} Witten, E.,  {\it Nucl. Phys.} {\tit B311}, 46 (1988/89).

\item{4.} Horne, J. H.,   {\it Nucl. Phys.} {\tit B334}, 669 (1990).

\item{5.} Nelson, J. E., Regge,  T. and Zertuche, F..   {\it Nucl.
Phys.} {\tit B339} 516, (1990).

\item{6. } Koehler, K., Mansouri, F., Vaz, C. and Witten, L.,  {\it
Mod. Phys. Lett.} {\tit A5}, 935 (1990);  {\it Nucl. Phys.} {\tit
B341}, 167 (1990) and  {\it Nucl. Phys.} {\tit B348}, 373 (1991).

\item{7. } Urrutia, L., F., Waelbroeck, H.  and Zertuche, F.,  {\it Mod.
Phys. Lett.} {\tit A7}, 2715 (1992).

\item{8. }  Mandelstam, S.,  {\it Phys. Rev.} {\tit D}, 2391 (1979).

\item{9. }Berenstein, D., E. and Urrutia,  L., F.,  Preprint   
Instituto de
Ciencias Nucleares ICN-UNAM-17-93, may 1993.

\item{10. } For a review see for example Loll., R., {\it Teor. Mat.  
Fiz.} {\bf
93}, 481 (1992) and  Preprint  Pennsylvania State University CGPG-93/9-1,
september 1993.
\item{} Br\"{u}gmann, B., Preprint Max Planck Institute of Physics,
MPI-Ph/93-94, december 1993.

\item{11. } For a  discussion of this point see for example  
Henneaux, M. and
Teitelboim, C., {\it Quantization of Gauge Systems}, Princeton University
Press, New Jersey, 1993,  Chapter 6.

\item{12. } See for example De Witt, B.,  {\it Supermanifolds},
Cambridge University Press, Cambridge,1984.

\item{13. } Backhouse, N., B. and Fellouris, A., G.,  {\it J. Phys.  
}{\tit
A17}, 1389 (1984).

\item{14. } Kobayashi, Y. and Nagamachi, S.,  {\it J. Math. Phys.}
{\tit 31},  2726 (1990).

\item{15. } See for example Nering, E., D.,  {\it Linear Algebra and
Matrix Theory}, Second edition, John Wiley, New York, 1970.

\item{16. } Urrutia, L., F. and Morales, N.,  {\it J. Phys. }{\tit  
A26}, L441
(1993).
\item{}Urrutia, L., F., The relation between the Mandelstam and the
Cayley-Hamilton identities: their extension to the case of  
supermatrices, to
appear in  the {\it Proceedings of SILARG-8}, edited by Letelier,  
P., et. al.,
World Scientific.
\item{}Urrutia, L., F. and Morales, N., The Cayley-Hamilton theorem for
supermatrices, to appear in the proceedings {\it Aspects of General  
Relativity
and Mathematical Physics}, dedicated to the 65-th birthday of Jerzy
Pleba\`{n}sky, edited by N.  Bret\'on, R. Capovilla  and  T. Matos.

\item{17.}Urrutia, L., F. and Morales, N., {\it J. Phys. }{\tit  
A27}, 1981
(1994).

\end